\documentclass[12pt]{article}
\usepackage{enumerate}
\usepackage[hyperindex,breaklinks,colorlinks,citecolor=blue,pagebackref]{hyperref}
\usepackage[latin1]{inputenc}
\usepackage{amsmath, amsthm, amsfonts, amssymb}
\usepackage{mathrsfs}
\usepackage{latexsym}
\usepackage{fullpage}
\usepackage[all]{xy}
\usepackage{color}
\usepackage{stmaryrd}

\newtheorem{theorem}{Theorem}[section]
\newtheorem{corollary}[theorem]{Corollary}

\newtheorem{lemma}[theorem]{Lemma}

\theoremstyle{definition}

\newtheorem{remark}[theorem]{Remark}

\newcommand{\Sp}{\ensuremath{S}}
\newcommand{\SSp}{\ensuremath{\mathbb{S}}}

\newcommand{\fs}{\{0,1\}^*}
\newcommand{\N}{\mathbb{N}}
\newcommand{\NN}{\mathcal{N}}
\newcommand{\tuple}[1]{\langle #1 \rangle}
\newcommand{\lfs}{\mathsf{LFS}}
\newcommand{\lfsst}{\lfs^*}
\newcommand{\LFS}{\mathbf{LFS}}
\newcommand{\LFSst}{\mathbf{LFS}^*}
\newcommand{\cs}{\{0,1\}^\omega}
\newcommand{\zero}{\mathbf{0}}
\newcommand{\ptime}{\mathsf{P}}
\newcommand{\nptime}{\mathsf{NP}}
\newcommand{\exptime}{\mathsf{EXPTIME}}
\newcommand{\bpp}{\mathsf{BPP}}
\renewcommand{\L}{\mathcal{L}}
\renewcommand{\P}{\mathcal{P}}

\newcommand{\comp}{\mathsf{COMP}}
\newcommand{\pspace}{\mathsf{PSPACE}}
\newcommand{\uh}{\upharpoonright}

\newtheorem{proposition}[theorem]{Proposition}

\begin{document}

\title{On Low for Speed Oracles}

\author{Laurent Bienvenu \\
    LIRMM\\CNRS \& Universit\'e de Montpellier, France\\
     \and
  Rod Downey\\
 School of Mathematics and Statistics\\
  Victoria University of Wellington, New Zealand
}


\maketitle

\begin{abstract}
Relativizing computations of Turing machines to an oracle is a central concept in the theory of computation, both in complexity theory and in computability theory(!). Inspired by lowness notions from computability theory, Allender introduced the concept of ``low for speed'' oracles. An oracle~$A$ is low for speed if relativizing to~$A$ has essentially no effect on computational complexity, meaning that if a decidable language can be decided in time $f(n)$ with access to oracle~$A$, then it can be decided in time $poly(f(n))$ without any oracle. The existence of non-computable such $A$'s was later proven by Bayer and Slaman, who even constructed a computably enumerable one, and exhibited a number of properties of these oracles as well as interesting connections with computability theory. In this paper, we pursue this line of research, answering the questions left by Bayer and Slaman and give further evidence that the structure of  the class of low for speed oracles is a very rich one. 
 \end{abstract}

\section{Introduction}

The subject of this paper is oracle computation, more specifically 
the effect of oracles on the speed of computation. 
There are many notable results 
about oracles in classical complexity, beginning with the 
Baker-Gill-Solovay result \cite{BakerGS1975} which asserts that there are oracles~$A$ such that $\ptime^A=\nptime^A$, but that there are also oracles~$B$ such that $\ptime^B \not= \nptime^B$ (thus demonstrating that methods that relativize will not suffice to solve basic questions like $\ptime$ vs $\nptime$).
An underlying question is whether oracle results can say things about complexity questions in the unrelativized world.
Eric Allender and his co-authors 
\cite{AllenderBK2006,AllenderFG2013} showed that 
oracle access to the sets of random strings 
could give insight into 
basic complexity questions. 
For example, in~\cite{AllenderFG2013}, Allender et. al.
showed that 
$\cap_U \ptime^{R_{K_U}}\cap \comp\subseteq \pspace$
where $R_{K_U}$ denotes the strings whose prefix-free Kolmogorov complexity
(relative to universal machine $U$) is at least their length, and 
$\comp$ denotes the collection of computable sets.
Later the ``$\cap \comp$'' was removed by 
Cai et. al. \cite{CaiDELM2014}.
Thus we conclude that reductions to very complex sets like the 
random strings somehow gives insight into 
very simple things like computable sets.

Inspired by lowness notions in computability theory, Allender asked whether 
there were non-trivial sets which were ``low for speed'' in that, 
as oracles, they did not accelerate 
running times 
of computations by more than a polynomial amount.
Of course, as stated this makes little sense since 
using any $X$ as oracle, we can decide membership in~$X$ in linear time, while without oracle $X$ may not even be computable at all!
Thus, what we are really interested in is the set of oracles which do not speed-up the computation of \emph{computable sets} by more than a polynomial amount. More precisely, an oracle~$X$ is \emph{low for speed} if for any computable language~$L$, if some Turing machine $M$ with access to oracle~$X$ decides $L$ in time~$f$, then there is a Turing machine~$M'$ without oracle and polynomial~$p$ such that $M'$ decides~$L$ in time~$p \circ f$. (Here computation time of oracle computation is counted in the usual complexity-theoretic fashion: we have a ``query tape'' on which we can write strings, and once a string~$x$ is written on this tape, we get to ask the oracle whether $x$ belongs to it in time $O(1)$). 

There are trivial examples of such sets, namely oracles that belong to~$\ptime$, because any query to such an oracle can be replaced by a polynomial-time computation. Allender's precise question was therefore: 

\begin{center}
Is there an oracle $X \notin \ptime$ which is low for speed?
\end{center}

Such an~$X$, if it exists, has to be non-computable, for the same reason as above (if $X$ is computable and low for speed, then $X$ is decidable in linear time using oracle~$X$, thus -- by lowness -- decidable in polynomial time without oracle, i.e., $X \in \ptime$). 
 
A partial answer was given by Lance Fortnow (unpublished), who observed the following.

\begin{theorem}[Fortnow] If $X$ is a hypersimple and computably enumerable oracle, then $X$ is low for polynomial time, in that if $L \in \ptime^X$, then $L \in \ptime$.
\end{theorem}

Allender's question was finally solved by Bayer and Slaman, who showed the following. 

\begin{theorem}[Bayer-Slaman \cite{Bayer-PhD}] 
There are non-computable, computably enumerable, sets~$X$ which are low for speed.  
\end{theorem} 

Once their existence is established, it is natural to wonder what kind of sets might be low for speed. A precise characterization seems currently out of reach, but it is interesting to see how lowness for speed interacts with other computability-theoretic properties. One needs however to keep in mind that lowness for speed is \emph{not} closed under Turing equivalence: as we saw above that in the $\mathbf{0}$ degree (computable sets) some members are low for speed and others that are not (on the other hand it is easy to see that if $A$ is polynomial-time reducible to $B$ and $B$ is low for speed, then $A$ is also low for speed). 

In his PhD thesis, Bayer showed that if $X$ is computably enumerable and of promptly simple Turing degree, then $X$ is \emph{not} low for speed, but also proved that this did not characterize the computable enumerable oracles that are low for speed. Bayer also studied the size of the set of low for speed oracles, where `size' is understood in terms of Baire category. Surprisingly, whether the set of low for speed oracles is meager or co-meager depends on the answer of the famous $\ptime =? \nptime$ question. 

In this paper, we continue Bayer and Slaman's investigation on the set of low for speed oracles. In the next section, we give an easier proof of the existence of non-computable low for speed oracles which does not require the full Bayer-Slaman machinery (but the oracle we construct is not computably enumerable). In Section~\ref{sec:ce-lfs}, we focus on the computably enumerable low for speed oracles, and prove that -- quite surprisingly -- they cannot be low in the computability-theoretic sense, but can however be low$_2$. Finally, we pursue Bayer and Slaman's idea to study how large the set of low for speed oracles is, in terms of measure and category. In particular, we solve a question they left open by showing that the set of low for speed oracles has measure~$0$ and obtain some interesting connections with algorithmic randomness. Finally, though lowness for speed is not closed under Turing equivalence
it is nonetheless natural to ask which Turing degrees contain a low for speed member, which is what Section~\ref{sec:lfs-T-degrees} is about. 

Throughout this paper, we will denote by $\fs$ the set of finite strings. In our setting, an oracle is a \emph{language}, i.e., a subset of $\fs$; however, as is typical in computability theory, it is more convenient in some of the results we present below to view oracles as infinite binary sequences (whose set we denote by $\cs$), by first identifying finite strings with integers (the $(n+1)$-th string in the length-lexicographic order being identified with~$n$) making the oracle a subset of $\N$ (which is what we mean when we talk about a `set' without further precision) and then identifying the oracle with its characteristic sequence (the $(n+1)$-th bit is $1$ if~$n$ belongs to the oracle, $0$ otherwise. When building oracles~$X$ with certain computability-theoretic properties, viewed as infinite binary sequences, we will often need to refer to prefixes of~$X$, which are themselves binary strings. To avoid confusion between \emph{members} and \emph{prefixes} of oracles, we will use latin letters $x, y, z, \ldots$ to denote members of oracles, and greek letters $\sigma, \tau, \ldots$ for prefixes of oracles. Two strings $\sigma$ and $\tau$ are \emph{incompatible} if for some $i < \min(|\sigma|,|\tau|)$, $\sigma(i) \not= \tau(i)$. We denote this by $\sigma \, \bot \, \tau$. The join $X \oplus Y$ of two infinite binary sequences $X, Y$ is the sequence $X(0)Y(0)X(1)Y(1) \ldots$. Finally $X \uh n$ is the prefix of~$X$ of length~$n$. 

Our paper requires some knowledge of computability theory and algorithmic randomness. One can consult the book~\cite{DowneyH2010} for the results and concepts we allude to below. Our notation is mostly standard.  We denote Cantor's pairing function by $\langle .,. \rangle$. We also fix an effective list $(\Phi_e)$ of all oracle Turing functionals (or machines: $\Phi_e^A$ is the Turing machine of index $e$ with oracle~$A$, which for a fixed~$A$ is a partial function from $\fs$ to $\{0,1\}$). For a given functional $\Phi_e$ and oracle~$A$, $time(\Phi_e^A,x)$ denotes the running time of $\Phi_e$ on input~$x$ with oracle~$A$ (counting time according to the model of computation described above) and $time(\Phi_e^A)$ is the function $x \mapsto time(\Phi_e^A,x)$. We let $(R_i)$ be an effective enumeration of all partial computable functions from $\fs$ to $\{0,1\}$. We denote the set of low for speed oracles by $\lfs$, and the subset of $\lfs$ consisting of its non-computable elements by $\lfsst$.

\section{Existence of non-computable low for speed sets}

In this section we will present a simple proof of the existence of a non-computable low for speed oracle. Define the set $\Sp$ of strings by $\Sp = \{0^{2^n} \mid n \in \N\}$ 
and let $\SSp$ be the set of `sparse' sets of strings which only contain strings from $\Sp$, that is, $\SSp = \{X \in \cs  \mid  X \subseteq \Sp\}$. 

By extension, we say that a string $\sigma$ is in $\SSp$ if it is a prefix of some element of $\SSp$. 
The interest of the set $\SSp$ is that there are only $O(n)$ strings in $\SSp$ of length~$n$. Thus, given a Turing machine~$\Phi$, it is possible to simulate in time $poly(t)$ the behaviour of $\Phi^X$ during~$t$ steps of computation on all~$X \in \SSp$ (an idea which is already present in the Bayer-Slaman argument presented in the next section). 

\begin{theorem}\label{thm:existence}
There exists a non-computable~$X$ which is low for speed. 
\end{theorem}

\begin{proof}

We want~$X$ to satisfy all requirements $\mathcal{R}_{(e,i)}$, where $e,i$ range over integers, defined as follows\\

$\mathcal{R}_{(e,i)}$: either $R_i$ is partial, or $\Phi_e^X \not= R_i$, or  $\Phi_e^X = R_i$ but the computation of $R_i$ via $\Phi^X_e$ can be simulated by a functional $\Psi$ running in time polynomial in $time(\Phi^X_e)$.\\

We build our oracle~$X$ by finite extension. Let $\sigma_0$ be the empty string. At stage $s+1=\tuple{e,i}$, do the following. 
\begin{itemize}
\item[(a)] If there is an $n$ and a $\tau \in \SSp$ extending $\sigma_s$ such that $\Phi_e^\tau(n)$ and $R_i(n)$ both converge and have different values, then let $\sigma_{s+1}$ be the first (say in length-lexicographic order) such string~$\tau$. 
\item[(b)] If there is no such string~$\tau$, then take $\sigma_{s+1} = \sigma_s 0$
\end{itemize}

Finally let~$X$ be the unique infinite sequence extending all~$\sigma_s$. We claim that~$X$ is as wanted. Let us first prove that~$X$ must be incomputable. Suppose $X=R_i$ for a total $R_i$. Let $e$ be an index such that $\Phi_e$ is the identity functional. By construction, when choosing the prefix $\tau$ of~$X$ at stage~$s+1=\tuple{e,i}$, we must be in case (a), and thus $\tau$ is precisely chosen to ensure $X \not= R_i$, a contradiction. Let us now prove that $X$ is low for speed. Fix a pair $(e,i)$ let $s+1=\tuple{e,i}$, and let us see how $\sigma_{s+1}$ was constructed. If we were in case (a) at that stage, we have ensured $\Phi_e^{\sigma_{s+1}} \, \bot\, R_i$ and thus $\Phi_e^X \, \bot \,R_i$, thereby satisfying $\mathcal{R}_{(e,i)}$. If we were in case~(b), there are three subcases:
\begin{itemize}
\item Either $R_i$ is partial, then the requirement $\mathcal{R}_{(e,i)}$ is satisfied
\item Or there is an~$n$ such that $\Phi_e^\tau(n) \uparrow$ for any extension $\tau$ of $\sigma_s$, in which case $\Phi_e^X(n) \uparrow$ and thus $\Phi_e^X \not= R_i$ should $R_i$ be total. 
\item Or, if we are in neither of the two above cases, for every $n$ there is an extension $\tau$ of $\sigma_s$ such that $\Phi_e^\tau(n) \downarrow$, and for any such $\tau$, we have $\Phi_e^\tau(n) = R_i(n)$. In this case, we can build a functional $\Psi$ which computes $R_i$ as follows. On input~$n$, at stage $t$, it computes $\Phi_e^\tau(n)$ during~$t$ steps of computation for all~$\tau \in \SSp$ of length~$t$ extending $\sigma_s$. If a $\tau$ is found such that  $\Phi_e^\tau(n) \downarrow$, then we set $\Psi(n)=\Phi_e^\tau(n)$. As we already mentioned, there are only~$O(t)$ strings of length~$t$ in $\SSp$ and it is obvious that they can be listed in polynomial time. Hence, simulating all computations $\Phi_e^\tau(n)$ during~$t$ steps can be done in time $p(t)$ for some polynomial~$t$. This shows that for any $Y \in \SSp$ extending $\sigma_s$, if $\Phi_e^Y(n)$ returns (the value of $R_i(n)$) in time $t$, this is found out by the procedure $\Psi$ at stage~$t$, which corresponds to $\sum_{s \leq t} p(s) +O(1)$ steps of computation for~$\Psi$, which is also polynomial in~$t$. This being true for any~$Y \in \SSp$ extending $\sigma_s$, we have in particular that $time(\Psi) = poly(time(\Phi_e^X))$. 
\end{itemize}
\end{proof}

One should note that the case disjunction in this proof is a $\Sigma_1$/$\Pi_1$ dichotomy, and therefore one can carry out the construction below $\zero'$, therefore establishing the existence of a $\zero'$-computable set that is low for speed. This is weaker than the Bayer-Slaman result presented in the next section, which asserts the existence of a \emph{c.e.} such set. However, this proof is both simpler and, as we will see in the remainder of the paper, has further useful corollaries. \\

\section{Computably enumerable low for speed sets}\label{sec:ce-lfs}

We now restrict ourselves to the computably enumerable (c.e.) sets, and study which of these can be low for speed. For the sake of completeness, we present the main ideas of the proof of Bayer and Slaman~\cite{Bayer-PhD} that there are indeed c.e.\ sets in $\lfsst$.

\begin{theorem}[Bayer-Slaman Theorem]
There exist c.e.\ non-computable sets that are low for speed.
\end{theorem}

\begin{proof}[Proof sketch]
The proof uses a tree-of-strategies argument. We need to satisfy
\[
\P_e: \overline{A}\ne W_e,
\]
and
\[
\L_{e,i}:  \mbox{If }\Phi^A_e =R_i\mbox{ total, then some }
\Psi \mbox{ computes }
R_i \mbox{ in time polynomial in }time(\Phi_e^A).
\]
The $\P_e$-strategy is a standard Friedberg-Muchnik strategy on a  tree. 
A node~$\rho$ devoted to this requirement picks a fresh follower 
$x$, waits for $x\in W_{e}[s]$ and if this happens puts $x$ into~$A$. 

The basic strategy for $\L_{e,i}$ is the following. First, throughout the whole construction of~$A$, we will promise that if we add an element~$x$ to~$A$ at stage~$t$, then we must immediately also add all $y \in [x,t]$ (this is often referred to as a  \emph{dump construction}). This way, at any stage~$s$, there will only be at most~$s$ strings $\alpha$ of length~$s$ that can potentially be a prefix of (the final)~$A$. And thus -- just like in the previous section -- at stage~$s$, it is possible to emulate all computations $\Phi_e^\alpha(x)[s]$ for all such $\alpha$'s and $x \leq s$ in time $poly(s)$. 

When the strategy is eligible to act at stage~$s$, for every $x \leq s$ on which $\Psi$ is not defined yet, it computes all $\Phi_e^\alpha(x)[s]$ for all potential prefixes $\alpha$ of~$A$, and should one of them converge, defines $\Psi(x)$ to be the value of $\Phi_e^\alpha(x)$ for the $\alpha$ that has the fastest convergence. If no $\Phi_e^\alpha(x)[s]$ converges, $\Psi(x)$ remains undefined until the strategy is eligible to act again. 

Now, if at some later stage we find a value $x$ such that $R_i(x) \downarrow$ and $\Psi(x) \not= R_i(x)$, then we find the $\alpha$ such that $\Psi(x) = \Phi_e^\alpha$ and add elements into~$A$ so that $\alpha$ becomes a prefix of $A$. This ensures $\Phi_e^A \not= R_i$ and terminates the strategy. All strategies of lower priorities are then injured and must be reset. If we never find such an~$x$, this means that either $\Phi_e^A$ is partial, or $R_i$ is, or $\Psi = \Phi_e^A = R_i$ and by construction the running time of $\Psi$ is polynomial in the running time of $\Phi_e^A$.\footnote{Actually, there is a subtlety here: one must ensure that the strategy for $\L_{e,i}$ is eligible to act often enough, i.e., allowed to act for the $n$-th time before stage~$q(n)$ for some polynomial~$q$, but this can easily be ensured.}

This is enough to ensure the success of the strategy in isolation. The difficulty comes from the interaction with lower-priority strategies which might want to add elements into~$A$. The final key to the Bayer-Slaman proof is the following. Suppose that at some stage $s$ a strategy of lower priority wants to add an interval $[y,s]$ of elements into~$A$. The problem is that 
the computations on this configuration might be \emph{slow}. Perhaps for some 
$x$ of length $\leq s$ we have not as yet seen 
$\Phi_e^{A_s\cup [y,s]}(x)\downarrow$. Even more importantly, we don't even know 
that the value of this will agree with the value $\Psi(x)$ we have already defined.

The idea is the following. $R_i$ has to confirm the computations, that is, we wait until $R_i(x)$ converges on all~$x$ where $\Psi$ has already been defined. When (and if) this happens, we must have $\Psi(x)=R_i(x)$ for all such~$x$ otherwise we would be in the above case where we can ensure $\Phi_e^A \not= R_i$ and satisfy the requirement. If this never happens, our requirement will be satisfied because $R_i$ would be partial. And if it does happen, then we can safely add $[y,s]$ to~$A$ because if this causes $\Phi_e^A(x)$ to change, it will yield $\Phi_e^A(x) \not= R_i(x)$ which satisfies the requirement. But there is one last problem: while waiting for this confirmation, the construction of $\Psi$ cannot wait as we need it to be as fast as $\Phi_e^A$. The crucial trick is, from the point of view of our strategy, to carry on \emph{as if} $[y,s]$ had already been enumerated into~$A$. Indeed, if the confirmation ever happens, the elements of $[y,s]$ will be truly enumerated into~$A$ which $\Psi$ will have correctly assumed ahead of time, and if the confirmation never happens, $\Psi$ might be wrong (i.e., $\Psi \not= \Phi_e^A$)  but this will not matter because in this case $R_i$ will be either partial or different from $\Phi_e^A$. 
Of course, when the confirmation never occurs the strategy of lower priority never gets to enumerate into $A$ the elements it wants. This is where we make use of a standard tree construction where strategies of lower priorities guess the outcome of strategies of higher priority. We refer the reader to~\cite{Bayer-PhD} for details.

\end{proof}

Within the c.e.\ sets, would one expect a set to be low for speed to 
have low or high information content?  In particular, one would expect that a low for speed c.e.\ set would 
be one with little computational power, in the same way that 
sets low for 1-randomness are all (super-)low (see Nies~\cite{Nies2009}). The next theorem is therefore quite surprising. 

\begin{theorem} \label{low} If $A$ is c.e.\ and of low Turing degree
(i.e. $A'\equiv_T \emptyset'$), then $A$ is 
\emph{not} low for speed.
\end{theorem}

\begin{proof}
Assume that $A$ is not computable, is c.e., and is low. Let $(\Phi_e,p_e)$ be an enumeration of pairs of one functional and one polynomial with coefficients in $\N$. We will build a Turing functional $\Psi$ and a computable set $R$ such that $\Psi^A=R$. This is our global requirement and we make the following global commitment: if a value $R(n)$ gets defined at some stage, $\Psi^X(n)$ is immediately defined to be equal to $R(n)$ for all $X$'s on which $\Psi^X(n)$ is still undefined. We want to satisfy, for each~$e$:
 \[
(\mathcal{R}_e): \text{$\Phi_e$ does not compute $R$ in time $p_e(time(\Psi^A))$}
\]
thus proving that $A$ is not low for speed. The strategy for a single requirement $(\mathcal{R}_e)$ is the following. Throughout the construction, we build a `verifier',  i.e., a partial computable~$S$ such that $S(e,.)$ is the attempt by the $(\mathcal{R}_e)$-strategy to guess~$A$. We also define an auxiliary functional $\Theta$ common to all strategies whose index we know in advance, and use the lowness of $A$ to obtain a computable 0-1 valued function $h(.,.)$ such that $\lim_t h(e,t)$ exists for all~$e$, and equals~$1$ when $\Theta^A(e) \downarrow$, $0$ otherwise. (Informally, $\Theta^X(e) \downarrow$ means that a prefix of~$X$ is believed to be a prefix of $A$ at some stage of the strategy for $(\mathcal{R}_e)$, and this will cause the strategy to enter Case 3 as described below.)

At the initial stage $s_1$, $S$ is empty and we pick a first fresh witness $w_1$ larger than any integer mentioned so far in the construction and define $\Psi^{A_{s_1} \uh 1}(w_1)=0$. Let $t_1$ be the time this computation takes. Now, check whether $\Phi_e(w_1)$ returns in $p_e(t_1)$ steps. We distinguish three cases:\\

\textbf{Case 1}: $\Phi_e(w_1)$ returns~$1$ in $\leq p_e(t_1)$ steps. In this case, we set $R(w_1)=0$ and $R(n)=0$ for all $n \leq w_1$ on which $R$ is still undefined, and commit to having $\Psi^A(w_1)=0$ even after potential future $A$-changes. This way we ensure $\Phi_e \not= R = \Psi^A$, thus immediately satisfying $(\mathcal{R}_e)$, and we stop the strategy for this requirement. \\

\textbf{Case 2}: $\Phi_e(w_1)$ returns~$0$ in $\leq p_e(t_1)$ steps. In this case, we do not define $R(w_1)$ just yet. Instead, we set $S(e,1)=A_{s_1} \uh 1$. We then create a second witness $w_2$ at stage $s_2$ and proceed as above for this new witness (with $A_{s_2} \uh 2$ in place of $A_{s_1} \uh 1$). And so on: for further occurrences of this case, the procedure will extend $S$ and create a witness $w_3$ at stage~$s_3$ looking at prefixes of length~$l=3$, etc (and if Case 2 then causes a reset, we stay at the same level $l$ when resetting). Meanwhile, we continue to monitor $A$. Again, there are two subcases for a given~$l$: 
\begin{itemize}
\item[(a)] At some point we discover that $A_{s_l} \uh i$ is not in fact an initial segment of $A$, we are then free to set $R(w_l)=1$ (which will guarantee $\Psi^A(w_l)=1 =R(w_l) \not= \Phi_e(w_l)$ since we only committed to $\Psi^{\sigma}(w_l)=0$ for $\sigma$'s that are not prefix of $A$), and this way we have satisfied $(\mathcal{R}_e)$. We then stop the strategy. 
\item[(b)] $A_{s_l} \uh l$ is a true initial segment of $A$, in which case nothing further will happen regarding witness~$w_l$. What is gained is that $S(e,l)$ will be defined to be $A_{s_l} \uh l = A \uh l$, thus progress was made towards computing~$A$.\\
\end{itemize}

\textbf{Case 3}: $\Phi_e(w_1)$ is still undefined after $p_e(t_1)$ steps. In this case, we set $\Theta^{A_{s_1} \uh 1}(e) \downarrow$ (which should be interpreted as signalling that we have entered Case 3). Observe that is $A_{s_1} \uh 1$ is a true prefix of~$A$, this implies $\Theta^{A}(e) \downarrow$ and therefore we would have $\lim h(e,t) = 1$. We distinguish two subcases

\begin{itemize}
\item[(a)] The current value $h(e,s)$ is~$0$. Then we wait for a stage $t>s$ such that either $h(e,t)=1$ or  $A_{t} \uh 1 \not= A_s \uh 1$ (one of the two must happen as we explained above). If the former happens first we move to subcase (b) below. If the latter happens first, we restart the procedure resetting $s_1$ to the current stage~$t$ and keeping the same $w_1$. 
\item[(b)] The current value $h(e,s)$ is~$1$. We then set $R(w_1)=0$, set $R(n)=0$ for all $n \leq w_1$ on which~$R$ is still undefined and terminate the strategy \emph{for now}. However, if at a later time $t>s$, we see that $h(e,t)=0$ \emph{and} $A_t \uh 1 \not= A_{s_1} \uh 1$, then we resurrect the strategy and start over at the level~$i$ where we left off. 
\end{itemize}

We claim that this strategy satisfies the requirement $(\mathcal{R}_e)$. If Case 1 happens for any witness~$w_l$, the requirement is satisfied. Case 3a can only happen finitely many times at a given level since as $A_{s_l} \uh l$ can only change finitely many times. Case 3b can only happen finitely many times across all levels as each passage through this case causes a flip of $h(e,.)$, and we know $h(e,.)$ converges. Case 2b can also happen only finitely often, because each time we go through this case and do not get to diagonalize, $S(e,.)$ computes a longer initial segment of~$A$, but $A$ is incomputable so $S(e,l)$ would eventually have to be wrong.

Thus we either eventually end up in Case 1 (and immediately succeed) or Case 2a (and immediately succeed) or a terminal Case 3b, i.e., the strategy enters Case 3b and stays there forever. It remains to check this last scenario. Suppose the terminal Case 3b happens for some $A_{s_l} \uh l$ which is not a prefix of~$A$, this means that $\Psi^{A}(w_l)$ has not been defined yet and thus, should nothing else happen, we would have $\lim_t h(e,t)=0$ and would see a change in $A \uh l$, thus leaving this occurrence of Case 3b, a contradiction. So $A_{s_l} \uh l$ is indeed a prefix of~$A$ and by construction $\Psi^A(w_l)$ returns $0=R(i)$ in a number of steps~$t$ while $\Phi_e(w_l)$ does not return in less than $p_e(t)$ steps, thus the requirement is satisfied. It is now straightforward to satisfy all requirements by ordering them in order of priority, noticing that each strategy only makes finitely many changes to $R$ before achieving its goal and $R$ is total as every time $R(w)$ becomes defined, so do the $R(n)$ for $n \leq w$ that were previously undefined.

\end{proof}

It is important to note that the above result fails to hold outside of the c.e.\ setting. 

\begin{theorem}\label{thm:low-lfs}
There exists a low, non-computable set $X$ which is low for speed. 
\end{theorem}

\begin{proof}
See Section~\ref{sec:lfs-T-degrees}.
\end{proof}

We next ask whether Theorem~\ref{thm:low-lfs} is tight in terms of the 
lowness/highness hierarchy. Recall that 
$A$ is low$_2$ if 
$A''\equiv_T \emptyset''$.

\begin{theorem} \label{thm:lfs-low2}
There is a low$_2$ c.e.\ set that is low for speed.
\end{theorem}

\begin{proof}
This time we must build $A$ to satisfy additionally to $\P_e$ and $\L_e$:
\[
\NN_e: (\Phi_e^A \text{~is total}) \to (\Theta_e \text{~is~total}).
\]
where $(\Theta_e)$ is a sequence of functionals we build.\\

The basic idea is that at stages where the length 
of convergence 
$\ell(e,s)=\max\{x\mid (\forall y \le x)\  \Phi_e^{A}(y)[s]\downarrow\}$ 
increases we would like to preserve this computation and set $\Theta_e(x) \downarrow$. We think of this having subrequirements 
$\NN_{e,x}$ devoted to preserving $\Phi_e^A(x)[s]$.

There is an apparent conflict between our requirements. Suppose we have three requirements $\NN_i$, $\L_j$, $\P_k$, in decreasing order of priority, and $\P_k$ believes that $\L_j$ will have infinitary outcome. At some point, $\P_k$ wants to enumerate some element $x$ into~$A$. As we saw in the previous construction, $\L_j$ will start behaving as if $x$ had already entered, while $\P_k$ waits for confirmation before actually doing so. However, if at some point in the process $\NN_i$ imposes a restraint containing~$x$, this will prevent $x$ from entering~$A$ and thus destroys $\L_j$'s plan to add~$x$ after confirmation, and the functional $\Psi$ built by $\L_j$ might just be wrong. One can check that any other interaction between requirements does not cause any conflict. 

The solution to the problematic case is to look closer at the tree of strategies. If the strategies for $\L_j$ and $\P_k$ are working under the assumption that $\NN_i$ has finitary outcome (waits forever to preserve $\Phi_e^A \uh x$ for some~$x$), there is no problem: if it is indeed the case they will not be bothered by any restraint $\NN_i$ might impose, and if they are wrong they will not be on the true path anyways. So the problematic case only happens if they are working under the assumption that $\NN_i$ has infinitary outcome (arbitrarily long restraints). The idea is, under the assumption that the requirement $\NN_i$ has infinitary outcome, to break it down to subrequirements $\NN_{i,x}$ devoted to preserve $\Phi_i^A \uh x$) and place them below on the tree, interleaved with other strategies. So now we are down to the case where $\L_j$ and $\P_k$ are working below an $\NN_{i,x}$, under the assumption that $\NN_i$ has infinitary outcome. But then there is no problem at all: because of the infinitary assumption, the strategies for  $\L_j$ and $\P_k$ can both wait for $\NN_{i,x}$ to place its restraint, and then only allow bigger elements to enter~$A$, thus respecting this restraint. 

There is no other difficulty to take care of when putting the strategies on a tree, except a small detail: when an $\NN$-strategy operate below an $\L$-strategy assuming infinitary outcome, when the $\L$ strategy behaves as if some $x$ has already entered~$A$, then so should the $\NN$-strategy. The rest is routine. 
\end{proof}

We can also combine the  the same ideas (dump construction together with awaiting for certification) with the standard proof that there exists an incomplete c.e.\ set~$A$ of high Turing degree (i.e., $A' \equiv_T \emptyset''$) to get the following. 

\begin{theorem} 
There is a high c.e.\ set $A$ which is low for speed.
\end{theorem}

Note that we do not say in the theorem that $A$ is Turing incomplete, because this in fact follows from lowness for speed, as we will see in Proposition~\ref{prop:lfsst-nullset}. 

\section{How big is $\lfs$?}

Bayer and Slaman showed that whether $\lfs$ is meager or not... depends on the answer to $\ptime$ vs $\nptime$ question! More precisely, if $\ptime=\nptime$, then $\lfs$ is co-meager (more precisely, every $2$-generic is low for speed), while if $\ptime\not=\nptime$, then $\lfs$ is meager (more precisely, every recursively generic is not low for speed). They left as an open question whether $\lfs$ has measure (Lebesgue) $0$ or $1$ (by Kolmogorov's $0/1$-law, it has to be one or the other). One might expect that, just like the meagerness of $\lfs$ depends on the  $\ptime$ vs $\nptime$ question, its measure depends on complexity-theoretic assumptions, such as the `$\ptime$ vs $\bpp$' question. This is not the case: we show that $\lfs$ is -- unconditionally -- a nullset. 

\begin{theorem}\label{thm:lfs-nullset}
The set $\lfs$ has measure~$0$. Indeed, no Schnorr random is low for speed. 
\end{theorem}

\begin{proof}
We first build a computable set $R$ with the following properties: (1) the set $R$ contains at most one string of any given length and (2) for all~$e$, if $\Phi_e$ computes~$R$, then $time(\Phi_e,x) > 2^{|x|}$ for almost all~$x$. To do so, we declare at the beginning of the construction all indices~$e$ `active', and for every~$n$ in order, do the following. We compute $\Phi_e(x)$ during $2^{|x|}$ of computation for all $x$ of length~$n$ and all currently active $e \leq n$. If for some pair $(e,x)$ we see that $\Phi_e(x)$ converges, we take the smallest such pair in the lexicographic order (smallest $e$ first, then smallest~$x$), we diagonalize against $\Phi_e$ by setting $R(x) = 1-\Phi_e(x)$ (thus ensuring $\Phi_e \not= R$), as well as $R(y) =0$ for all $y \not= x$ of length~$n$, and then declares~$e$ inactive from now on. If no such pair $(e,x)$ exists, set $R(y)=0$ for all~$y$ of length~$n$. This finishes the construction of~$R$. It is clear that~$R$ is computable (it is even in $\exptime$) and that it contains at most one string of each length. To verify property (2), suppose $\Phi_e$ computes~$R$, and observe that for any $x$ such that $time(\Phi_e,x) \leq 2^{|x|}$, the only way~$\Phi_e$ can escape being diagonlalized against by~$R$ is when some $\Phi_i$ with $i<e$ is diagonlized against on strings of length~$n$, but this can only happen~$e$ times. Thus if $\Phi_e$ does indeed compute~$R$, it must do so in time $2^{|x|}$ for almost all~$x$. 

Now we need to see how to speed up computations with a Schnorr random oracle. Consider the following procedure~$\Psi$. Given oracle~$Z$ and input~$x$, $\Psi^Z(x)$ first splits $Z$ (viewed as a binary sequence) as $Z=\zeta_1 \zeta_2 \ldots$ with $|\zeta_i|=i$ and $\Psi^Z(x)$ returns $0$ if $x=\zeta_{|x|}$ (thus the resulting computation is polynomial in $|x|$), and $\Psi^Z(x)=R(x)$ otherwise, using a fixed procedure to compute $R$. So there is a polynomial~$p$ such that for any $Z$, $time(\Psi^Z,x) \leq p(|x|)$ for infinitely many~$x$'s. Furthermore, we can only have $\Psi^Z(x) \not= R(x)$ if $x=\zeta_{|x|}$ and $\zeta_{|x|}$ happens to be the only string of its length in~$R$. This has probability at most $2^{-|x|}$ (`at most' because $R$ can also have no string of length~$|x|$ at all) if~$Z$ is chosen at random. This means that, by setting $C_n = \{Z \mid (\exists x)\,  |x| = n\, \wedge\, \Psi^Z(x) \not= R(x)\}$, we have $\lambda(C_n) \leq 2^{-n}$. 

The $C_n$'s are uniformly computable clopen subsets of $\cs$ because $\Psi$ is a tt-functional. Thus, a Schnorr random~$A$ can only belong to finitely many $C_n$'s (see for example~\cite[Lemma 1.5.9]{Bienvenu2008-phd}), meaning that $\Psi^A(x)=R(x)$ for almost all~$x$. Thus there is a finite variation $\hat{\Psi}$ of $\Psi$ such that $\hat{\Psi}^A=R$, and  $\hat{\Psi}^A(x)$ is computed in polynomial time for infinitely many~$x$ while $time(\Phi_i,x)>2^{|x|}$ for any~$\Phi_i$ computing~$R$ and almost all~$x$. This shows that $A$ is not low for speed. 
\end{proof}

On the other hand, we will see in the next section that $\lfs$ is large in the set-theoretic sense, namely that it has the size of the continuum. 

Finally, there is one last notion of size for subsets of $\cs$ that is dear to computability theorists, namely, a set is `large' if it contains a Turing upper cone and is `small' if it disjoint from a Turing upper cone. Martin's Turing determinacy theorem tells us that any Borel set which is closed under Turing equivalence must be either large or small on this account. The set $\lfs$ is indeed Borel (this is easy to see from the definition), but it is not closed under Turing equivalence, so Martin's theorem does not apply. In the next section, we will use a classical result from complexity theory to show that $\lfs$ is in fact disjoint from a Turing upper cone (Theorem~\ref{thm:dnc-not-lfs}).

\section{Lowness for speed and Turing degrees} \label{sec:lfs-T-degrees}


While lowness for speed is not closed under Turing equivalence, the following question is nonetheless interesting: \\

\emph{Which sets are Turing equivalent to some low for speed $X$? Which sets compute some non-computable low for speed $X$?}\\

We denote by $\LFS$ and $\LFSst$ the set of Turing degrees that contain a low for speed set and a non-computable low for speed set, respectively. One of the main results of Bayer~\cite{Bayer-PhD} is that not all degrees are in $\LFS$. Indeed, there exists a c.e.\ degree $\mathbf{a} \notin \LFS$. The main question left open by Bayer regarding $\LFS$ is whether it is downward closed under $\leq_T$ or closed under join. We give a negative answer to both questions. To show that it is not downward closed, we need the following extension of Theorem~\ref{low} to degrees. 

\begin{theorem}\label{thm:low-ce-degree}
For any low c.e.\ degree $\mathbf{a}>\mathbf{0}$, we have $\mathbf{a} \notin \LFS$. 
\end{theorem}

\begin{proof}
We now suppose that $A$ only is of c.e.\ degree (as opposed to being c.e.) and is low. Let $(A_s)$ be a $\Delta^0_2$ approximation of~$A$, and let $\mu$ be its modulus of convergence, i.e., $\mu(n)$ is the smallest~$s$ such that all $\{A_t \mid t \geq s\}$ have the same prefix of length~$n$ (which must thus be the prefix of $A$ of length~$n$). Observe that $\mu$ is a lower semi-computable function, and let $\mu_s$ be its approximation at stage~$n$ (setting $\mu_0(n)=0$ for all~$n$). 

The fact that $A$ has c.e.\ degree is equivalent to the fact that $\mu$ is $A$-computable (see~\cite{Soare2016}), so let $M$ be a functional such that $M^A(n)=\mu(n)$ for all~$n$. Consider the set $\{ X \in \cs \mid M^X(n)=k \}$. This is an effectively open set, uniformly in $(n,k)$ and thus can be represented by a c.e.\ set of strings that are pairwise incomparable for the prefix relation (by this, we mean that an $X$ is in $M^{-1}(n,k)$ if and only if one of these strings is a prefix of~$X$), and this representation is effective and uniform in $(n,k)$. We call $P(n,k)$ the set of strings which both belong to this set \emph{and} are a prefix of some approximation $A_s$ of $A$. Observe the following easy facts:
\begin{itemize}
\item[(i)] $P(n,k)$ is c.e.\ uniformly in $(n,k)$
\item[(ii)] $P(n,k)$ is finite for all $(n,k)$. 
\item[(iii)] If $k \not= k'$, then any two strings $\sigma \in P(n,k)$ and $\tau \in P(n,k')$ are pairwise incomparable for the prefix order.
\item[(iv)] $A$ has a prefix in $P(n,\mu(n))$ for all~$n$ (and thus no prefix in other $P(n,k)$ for $k \not= \mu(n)$ by the previous fact). 
\end{itemize}
Facts $(i)$, $(iii)$ and $(iv)$ follow directly from the definition. Fact $(ii)$ holds because the sequence $(A_s)$ pointwise converges. 

We now adapt the construction of $R$ and $\Psi$ as in the proof of Theorem~\ref{low}. The strategy to win against a pair $(\Phi_e,p_e)$ still works by levels, with a verifier $S$ which this time attempts to compute $\mu$ (it will eventually fail because if $\mu$ were computable, so would be~$A$), and again an auxiliary functional $\Theta$, with $h(.,.)$ a computable approximation of $\Theta^A$, by lowness of $A$. Also, as before, as soon as $R(n)$ becomes defined, $\Psi^X(n)$ is immediately set to be equal to $R(n)$ for any~$X$ on which $\Psi^X(n)$ was still undefined. Our strategy starts with $l=1$ and $S$ undefined everywhere, and does the following. First, it looks at the current value $k=\mu_s(l)$ and starts enumerating $P(l,k)$. When a new string $\sigma$ enters $P(l,k)$, we pick a fresh witness $w=w(l,\sigma)$ for this $\sigma$, set $\Psi^\sigma(w)=0$, look at the number $t$ of steps this computation takes, and compute $\Phi_e(w)$ during $p_e(t)$ steps of computation. Once again, there are three cases which will require different actions for this $\sigma$:\\

\textbf{Case 1}: $\Phi_e(w)$ returns~$1$ in $\leq p_e(t)$ steps. In this case, we set $R(w)=0$ and $R(n)=0$ for all $n$ below the current stage on which $R$ is still undefined, and stop the strategy forever.  \\

\textbf{Case 2}: $\Phi_e(w)$ returns~$0$ in $\leq p_e(t)$ steps. We set $S(e,l)=k$. We then start again the procedure with $l+1$ and 
 
\begin{itemize}
\item[(a)] If at some point we discover that $\mu(l)>k$, we are then free to set $R(w)=1$, $R(n)=0$ for all $n$ below the current stage on which $R$ is still undefined, and stop the strategy forever. 
\item[(b)] $\mu_s(l)=\mu(l)$, in which case $S(e,.)$  has made progress towards computing~$\mu$. 
\end{itemize}

\textbf{Case 3}: $\Phi_e(w)$ is still undefined after $p(t)$ steps. In this case, we set $\Theta^{\sigma}(e) \downarrow$. We have the same two subcases as before:

\begin{itemize}
\item[(a)] The current value $h(e,s)$ is~$0$ (here, this should be interpreted as the fact that $\emptyset'$ currently does not believe that any of the strings that have been enumerated so far in $P(l,k)$ and reached Case 3 is a true initial segment of~$A$). Then we wait for a stage $t>s$ such that either $h(e,t)=1$ or to see a new $\sigma$ enter $P(l,k)$, or to see an increase of $\mu$ so that $\mu(l)>k$. If $h(e,t)=1$ happens first, we move to subcase (b) below. If a new $\sigma$ enters $P(l,k)$, we set $R(w)=0$ and $R(n)=0$ for all  and run the strategy on the new $\sigma$. Finally, if $\mu(l)$ increases above~$k$, then we know that we have not seen any prefix of $A$ yet at this level, so we restart the procedure at the same level with the new value of~$k$. 
\item[(b)] The current value $h(e,s)$ is~$1$. We then set $R(w)=0$, set $R(n)=0$ for all $n$ below the current stage on which~$R$ is still undefined and terminate the strategy \emph{for now}. We resurrect the strategy at this level if we see either a new $\sigma$ entering~$P(l,k)$ or an increase in $\mu(l)$. 
\end{itemize}

On top of the action for a single~$\sigma$ that enters $P(l,k)$, if at any point of time we see a $\tau$ that enters $P(l,k)$, picks a witness $w_\tau$ and lands in Case 1 or Case 2, then we immediately stop the action of the other $\sigma$'s that we have seen so far in $P(l,k)$, and for all of those, copy the action of $\tau$ on witness $w_\tau$.  This is in fact only important in Case 2, the reason being that we do not want the action of a `bad' $\sigma$ (i.e., not a true prefix of~$A$) to interfere with the strategy on a good $\sigma$ that enters Case 2, as the bad $\sigma$ would otherwise (by going through Case 3 and defining $R(n)$ for the $n$ needed by the good $\sigma$). 

The rest of the verification is pretty much the same as before, with the additional point that we can only go through Case 3 finitely often at a given level because $h$ can only flip finitely many times \emph{and} $P(l,k)$ is a finite set, and go through Case 2b only finitely often otherwise $S(e,.)$ would be a computation of $\mu$.

\end{proof}


\begin{corollary}
$\LFS$ is not downward closed under $\leq_T$, even within c.e.\ degrees. 
\end{corollary}

\begin{proof}
Let~$\mathbf{a}>\mathbf{0}$ be a c.e.\ degree in $\LFS$ whose existence was explained in Section~\ref{sec:ce-lfs}. By Sacks's splitting theorem~\cite{Sacks1963b}, there is a low c.e.\ degree $\mathbf{0} < \mathbf{b} < \mathbf{a}$. By Theorem~\ref{thm:low-ce-degree}, $\mathbf{b} \notin \LFS$. 
\end{proof}

The next theorem will show that while not every c.e.\ degree contains a low for speed member, every non-zero c.e.\ degree $\mathbf{a}$ \emph{bounds} a degree $\mathbf{b} \in \LFS$. Recall Bayer's result that whether $2$-generics are low for speed or not depends on the `$\ptime$ vs $\nptime$' question. When it comes to the \emph{degree} of generics, we have that every $1$-generic is Turing-equivalent to a set that is low for speed, independently of complexity-theoretic assumptions. 

\begin{theorem}\label{thm:1gen-lfs}
Every 1-generic degree~$\mathbf{g}$ belongs to $\LFS$. 
\end{theorem}

\begin{proof}
We get this result by refining the proof of Theorem~\ref{thm:existence}. In that proof, we built an~$X$ low for speed by finite extension, and ensuring that $X$ was a subset of $\Sp= \{0^{2^n} \mid n \in \N\}$. For $G \subseteq \N$, let $\Sp_G= \{0^{2^n} \mid n \in G\}$. We claim that when $G$ is $1$-generic, $\Sp_G= \{0^{2^n} \mid n \in G\}$ is low for speed (and clearly $\Sp_G \equiv_T G$). In the proof of Theorem~\ref{thm:existence}, if we let $\mathcal{U_{e,i}}$ be the effectively open sets of those $Z$ such that for some~$n$, $\Phi_e^{S_Z}(n)$ and $R_i(n)$ both converge to different values, we know that $G$, being $1$-generic, is either in~$\mathcal{U}_{e,i}$ (hence satisfying the requirement $\mathcal{R}_{e,i}$ as per case (a)), or in the interior of the complement of~$\mathcal{U}_{e,i}$, which precisely corresponds to case (b), hence the requirement is also satisfied in this case. 
\end{proof}

We can derive a number of useful corollaries from this theorem. First of all, we see that~$\lfs$ has the size of the continuum since $G \mapsto S_G$ is one-to-one, and there are continuum many $1$-generic~$G$. We also get an immediate proof of Theorem~\ref{thm:low-lfs} that asserts the existence of a set of low degree that is low for speed. 

\begin{proof}[Proof of Theorem~\ref{thm:low-lfs}]
Take a $\Delta^0_2$ $1$-generic~$G$; the corresponding set $\Sp_G$ is low for speed and is low as it is both $\Delta^0_2$ and of $GL_1$ (Indeed a result of Jockusch~\cite{Jockusch1980} states that every $1$-generic degree~$G$ is $GL_1$, that is, $G' \equiv_T G \oplus \emptyset'$; when $G \leq_T \emptyset'$, this is equivalent to $G' \equiv_T \emptyset'$).
 \end{proof}

 A similar idea allows us to show that $\LFS$ contains a non-trivial interval in the Turing degrees.

\begin{corollary}\label{cor:segment-lfsst}
There is a degree $\mathbf{a} >\mathbf{0}$ such that every $\mathbf{0} \leq \mathbf{b}\leq\mathbf{a}$ is in $\LFS$.
 \end{corollary}

\begin{proof}
By a result of Haught~\cite{Haught1986}, if $\mathbf{a}$ is a $\Delta^0_2$ $1$-generic degree, every $\mathbf{b}>\mathbf{0}$ below $\mathbf{a}$ is of $1$-generic degree. Then the result follows immediately from Theorem~\ref{thm:1gen-lfs}.
\end{proof}

Another interesting corollary is that every non-computable c.e.\ set bounds a non-computable low for speed set. Likewise almost every set, in the measure-theoretic sense, \emph{bounds} a non-computable low for speed set. 

\begin{corollary}
Every non-zero c.e.\ degree bounds a member $\LFSst$, every $2$-random degree bounds a member of $\LFSst$. 
\end{corollary}

\begin{proof}
This is simply because every non-zero c.e.\ degree and every $2$-random degree bounds a $1$-generic degree~\cite{Kurtz1981,Kautz1991}. 
\end{proof}

\begin{theorem}
There are $\mathbf{a},\mathbf{b} \in \LFS$ such that $\mathbf{a} \vee \mathbf{b} \notin \LFS$
\end{theorem}

\begin{proof}
Let $G_0$ be $2$-generic, i.e., $1$-generic relative to $\emptyset'$. Consider $G_1 = G_0 \Delta \emptyset'$ where $\Delta$ is the symmetric difference. It is easy to check that $G_1$ is also $2$-generic. Thus $S_{G_0}$ and $S_{G_1}$ (defined as in the proof of Theorem~\ref{thm:1gen-lfs}) are both low for speed but $S_{G_0} \oplus S_{G_1} \geq_T G_0 \oplus G_1 \geq_T G_0 \Delta G_1 = \emptyset'$, so by the previous theorem, $deg(S_{G_0} \oplus S_{G_1}) \notin \LFS$. 
\end{proof}

After generic degrees, let us move to random degrees. We have seen in Theorem~\ref{thm:lfs-nullset} that $\lfs$ is a nullset, and in fact no Schnorr degree is low for speed. Interestingly, this result does not extend to Turing degrees: by a result of Nies et al.~\cite{NiesST2005}, every high degree has a Schnorr random member and we have seen that there is a low for speed of high degree. This leaves open the possibility that almost all~$X$ are \emph{Turing-equivalent} to a low for speed set. This would be similar to the category situation where -- under the reasonable assumption $\ptime \not= \nptime$ -- the set $\lfs$ is meager (as proven by Bayer and Slaman) but the set of $A$'s whose \emph{degree} is in $\LFS$ is co-meager (Theorem~\ref{thm:1gen-lfs}). This is not the case however: if we increase the algorithmic randomness level from Schnorr randomness to Martin-L\"of randomness, then the distinction disappears.

\begin{proposition} \label{prop:lfsst-nullset}
If $\mathbf{a}$ is, or simply bounds, a Martin-L\"of random degree then $\mathbf{a} \notin \LFS$ (equivalently: if $A \in \cs$ computes a Martin-L\"of random, $A$ is not low for speed). 
\end{proposition}

Note that this shows in particular that $\{A \mid deg(A) \in \LFS\}$ has measure~$0$, and also that any $A \geq_T \emptyset'$ is not low for speed, as Chaitin's $\Omega$ number is Martin-L\"of random and Turing equivalent to $\emptyset'$. 

Instead of proving Proposition~\ref{prop:lfsst-nullset} directly, we will prove the  following stronger theorem.

\begin{theorem}\label{thm:dnc-not-lfs}
If $\mathbf{a}$ is a DNC degree, then $\mathbf{a} \notin \LFS$ (equivalently: if $A \in \cs$ computes a DNC function, $A$ is not low for speed). 
\end{theorem}

Proposition~\ref{prop:lfsst-nullset} follows from this theorem because DNC degrees are closed upwards, and by a result of Ku\v cera~\cite{Kucera1985}, every Martin-L\"of random degree is a DNC degree. 

\begin{proof}
The proof of this theorem relies on the proof of a classical computational complexity theorem, namely Blum's speed-up theorem~\cite{Blum1971} (see also~\cite[Theorem 32.2]{Kozen2006}), which asserts that for every sufficiently fast growing computable function~$f$, there exists a computable set $R$ which admits no fastest algorithm in that for every~$i$ such that $\Phi_i=R$, there is a~$j$ such that $\Phi_j=R$ and $f(time(\Phi_j,x)) \leq time(\Phi_i,x)$ for almost every~$x$. 

Let us first discuss how to prove Blum's speed-up theorem. As it happens, we have already seen a proof with similar features, namely the proof of Theorem~\ref{thm:lfs-nullset} (this is no coincidence, as it is the proof Blum's speed-up theorem that inspired this other proof). We build~$R$ by diagonalization against all~$\Phi_i$, where for all~$x$ in order we try to find an active $i \leq |x|$ such that $\Phi_i(x)$ converges in less than $f^{|x|-i}(|x|)$ steps (here the exponent is understood as composition: $f^0$ is the identity, and $f^{n+1}=f \circ f^n$) and if such an~$i$ is found, we diagonalize against $\Phi_i$ by setting $R(x) = 1 - \Phi_i(x)$ for the smallest such~$i$, and declare~$i$ inactive from that point on. If no such~$i$ is found, set $R(x)=0$. Obviously $R$ is computable, and the same argument as in the proof of Theorem~\ref{thm:lfs-nullset} shows that for any $\Phi_i$ computing~$R$, one must have $time(\Phi_i,x) \geq f^{|x|-i}(|x|)$ for almost all~$x$. Now, suppose $\Phi_e$ is a functional that computes~$R$. We need to show that there is another functional which computes $R$ much faster than $\Phi_e$. Fix a large~$k$ and assume we are given as `advice' the finite list $\sigma_k$ of indices $i < k$ such that $\Phi_i$ eventually gets diagonalized against (and therefore~$i$ becomes inactive) in the construction of~$R$. Now, we can compute~$R$ via the following procedure. In a first phase, simply follow the construction of~$R$ as described above, until we reach a point where all $i \in \sigma_k$ have become inactive. At this point, we know (only because we know $\sigma_k$!) that none of the $\{\Phi_i \mid i  < k\}$ are relevant for the construction of~$R$ on future~$x$. Thus, we enter a second phase where to compute each~$R(x)$, we only need to simulate, for $k \leq j \leq |x|$, $\Phi_j(x)$ during $f^{|x|-j}(|x|)$ steps of computation. By dovetailing, this can be done in $poly(|x| \cdot f^{|x|-k}(|x|))$ (the polynomial being independent of~$k$) which, if $f$ is fast growing enough and~$k$ large enough compared to~$e$, is $< f \big(f^{|x|-e}(|x|)\big)$, which in turn is $< f(time(\Phi_e,x))$ for almost all~$x$ (note that such a $k$ can be computed uniformly given~$e$), and this finishes the proof of Blum's speed-up theorem. 

Looking more closely, the core of the argument is that for a given~$x$ and a given~$i$, if we somehow knew that $\Phi_i$ is not diagonalized against\emph{exactly} at that point~$x$, then we can save the computation of $\Phi_i(x)[f^{|x|-i}(|x|)]$ in the computation of~$R(x)$. We already see why DNC-ness naturally comes into play. Let $\theta : \N \rightarrow \fs$ be the partial computable function such that $\theta(i)$ is the one string~$x$ on which $\Phi_i$ is diagonalized against during the construction of~$R$, and $\theta(i) \uparrow$ if there is no such~$x$. Having a DNC function as oracle allows us to find, for any given~$i$, a $y \not= \theta(i)$ should $\theta(i)$ be defined, and thus one can speed up a bit the computation of $R(y)$ for this~$y$. In fact we can do much better: by a result of Jockusch~\cite{Jockusch1989}, having access to a DNC function allows us to uniformly avoid a finite number of values of a given partial recursive function. Here this means that with our DNC oracle, we can, uniformly in~$k$, find a~$y \notin \theta(i)$ for any $i<k$ on which~$\theta$ is defined, or equivalently, a $y$ such that none of the $\Phi_i$ with $i<k$ gets diagonalized at~$y$. Applying Blum's argument, one can indeed use this information to truly speed up the computation of $R(y)$. We are not quite done yet however: our argument shows that once we have computed~$y$ from~$k$ we can speed-up the computation of $R(y)$, but the computation of~$y$ itself might take a long time and offset the time we save on the computation of $R(y)$. 

To overcome this problem, we need to refine the above idea. Let $\tuple{.,.,.}$ be the canonical `tripling' function: $\tuple{a,b,c} = \tuple{a,\tuple{b,c}}$ and for $1 \leq i \leq 3$, let $\pi^i_3$ be the $i$-th projection ($\pi^i_3(\tuple{x_1,x_2,x_3})=x_i$, note that these function are polynomial-time computable). Since $A$ has DNC degree, again using Jockusch's result, $A$ computes, via a fixed functional $\Xi$, a function $F$ such that $F(k) \not= \pi^2_3(\theta(i))$ for any $i<k$ on which $\theta$ is defined (here and in the rest of the proof we identify strings and integers). Said otherwise, for \emph{any} pair $(a,c)$ of integers, none of  the functionals $\{\Phi_i \mid i<k\}$ gets diagonalized against at $y=\tuple{a,F(k),c}$. 

Now let $\Psi$ the functional which computes~$R$ as follows using oracle~$A$. On input~$x$, it first computes the projections of $x$, i.e., finds $k,l,m$ such that $x=\tuple{k,l,m}$. Then, it tries to compute $F(k)$ via $\Xi$ and using oracle~$A$ during~$m$ steps of computation. If it fails to do so, it then computes $R(x)$ using a fixed procedure (with no access to the oracle) and returns this value. If it does succeed to compute $F(k)$, it then checks whether $l=F(k)$. If not, $\Psi^A(x)$ again returns $R(x)$ using a fixed procedure to perform the computation, without access to the oracle. Finally, and this is the interesting case, if $F(k)$ is computed in $\leq m$ steps of computation, and $l=F(k)$, we know by definition of $F$ that none of the  $\{\Phi_i \mid i<k\}$ gets diagonalized against at $x$. Thus $\Psi^A$ can compute $R(x)$ by only simulating, like in the proof of Blum's theorem $\Phi_j(x)$ during $f^{|x|-j}(|x|)$ steps of computation for $k \leq j \leq |x|$. In that case, the total computation time is still $poly(|x| \cdot f^{|x|-k}(|x|))$ because the $m$ steps of computation needed to get $F(k)$ are simply $poly(|x|)$. For a given~$k$, taking $m$ sufficiently large will give enough time for the computation of $F(k)$ by $A$, and make $x=\tuple{k,F(k),m}$ large enough to ensure that the computation time of $\Psi^A(x)$ is $< f(time(\Phi_e,x))$, as long as $f$ is sufficiently fast growing. This shows that $A$ is not low for speed. 
\end{proof}

As an immediate corollary, we get the result mentioned in the pervious section that $\LFS$ is disjoint from a cone in the Turing degrees, again because DNC degrees are closed upwards. Another interesting consequence is that the analogue of the low basis theorem for $\Pi^0_1$ class does not extend to lowness for speed: it is not true that every non-empty $\Pi^0_1$ class contains a member of low for speed degree. For example, one can take a $\Pi^0_1$ class of Martin-L\"of randoms: all its elements have DNC degree by Ku\v cera's theorem, and thus do not have low for speed degree.

At this point, we know that the set of $X$'s which \emph{compute} a member of $\lfsst$ is very large: it has measure 1 and is co-meager, it contains every c.e.\ set, etc. We might even start thinking that \emph{every} non-computable $X$ computes a member of $\lfsst$. This is not the case however, as shown by the following theorem (which contrasts Corollary~\ref{cor:segment-lfsst}). 

\begin{theorem}\label{thm:minimal-bound-no-lfsst}
There is a degree $\mathbf{a}>\mathbf{0}$ such that no $\mathbf{0} < \mathbf{b} \leq \mathbf{a}$ is in $\LFSst$. Indeed, $\mathbf{a}$ can be chosen to be a minimal Turing degree. 
\end{theorem}

Note that another way to obtain a degree $\mathbf{a}>\mathbf{0}$ such that no $\mathbf{0} < \mathbf{b} \leq \mathbf{a}$ is in $\LFSst$ is by taking~$\mathbf{a}$ to be a hyperimmune-free degree containing a Martin-L\"of random member, which ensures that hat every $\mathbf{0} < \mathbf{b} \leq \mathbf{a}$ also computes a $1$-random (see~\cite[Corollary 8.6.2]{DowneyH2010}), and apply Proposition~\ref{prop:lfsst-nullset}. However, the existence of a minimal low for speed degree is interesting in its own right.

The classical construction of a minimal degree is done by forcing over total computable function trees (here we follow the terminology of~\cite{Soare2016}). A function tree is a partial function $T: \fs \rightarrow \fs$ such that if either $T(\sigma 0)$ or $T(\sigma 1)$ is defined, then all of $T(\sigma)$, $T(\sigma 0)$ and $T(\sigma 1)$ are, and $T(\sigma 0)$ and $T(\sigma 1)$ are strict extensions of $T(\sigma)$ such that $T(\sigma0) \, \bot \, T(\sigma1)$ (we say that $T(\sigma 0)$ and $T(\sigma 1)$ split $T(\sigma)$). We say that $\sigma$ is a node of $T$ if $\sigma \in rng(T)$. A tree $S$ is a sub-f-tree of $T$, which we denote by $S \preccurlyeq T$ when every node of $S$ is a node of~$T$. An infinite binary sequence~$Z$ is a path on an f-tree~$T$ if infinitely many prefixes of $Z$ are nodes of~$T$. The set of paths of $T$ is denoted by $[T]$. An f-tree naturally extends to a functional from $\cs$ to $\fs \cup \cs$ by setting $T(X) = \bigcup_{\sigma \preccurlyeq X} T(\sigma)$  When an f-tree~$T$ is total, this extension is an homeomorphism from $\cs$ to $\cs$, and if~$T$ is furthermore computable, its inverse $T^{-1}$ is also computable. Given a functional $\Phi_e$, we say that~$T$ is $e$-consistent if for any two nodes $\sigma$ and $\tau$ on~$T$ and any~$n$, if $\Phi_e^\sigma(n)$ and $\Phi_e^\tau(n)$ are both defined, then they are equal; in this case, for any path~$X$ of $T$, $\Phi_e^X$ is either partial or computable. We say that~$T$ is $e$-splitting if $T$ is total and for any $\sigma$, $\Phi_e^{T(\sigma 0)}$ and $\Phi_e^{T(\sigma 1)}$ are incomparable; in this case, the restriction of $\Phi_e$ to $[T]$ is total, one-to-one, and its inverse is computable.

%

The key lemma in the construction of a minimal degree states that for any total computable f-tree~$T$ and every~$e$, there is a total computable sub-f-tree~$S$ of~$T$ which is either $e$-consistent or $e$-splitting. Then an $X \in \cs$ of minimal degree is obtained by taking a sufficiently generic filter $G$ over the set of total computable f-trees ordered by $\preccurlyeq$, and take the intersection of their sets of paths (the non-computability of $X$ can be further ensured by remarking that for any $\sigma$, the set of computable f-trees whose nodes are all incomparable with $\sigma$ is a dense set for the order $\preccurlyeq$, thus one can choose~$X$ to avoid any fixed subset of $\cs$, such as its computable elements). 

We are going to prove Theorem~\ref{thm:minimal-bound-no-lfsst} by showing that taking a sufficiently generic filter for the order $\preccurlyeq$ ensures lowness for speed as well. For this, we will make use of the following lemma. 

\begin{lemma}
Let~$T$ be a total computable f-tree. There exists a total computable f-subtree $S \preccurlyeq T$ none of whose paths is low for speed. 
\end{lemma} 

\begin{proof}
The functional $T^{-1}: [T] \rightarrow \cs$ is total on its domain, which is a $\Pi^0_1$ class, and thus is a tt-reduction by effective compactness. Let $f$ be a computable time bound for the running time of $T^{-1}$, that is, for any $Y \in [T]$, whether $x \in T^{-1}(Y)$ can be decided in time $f(|x|)$ with access to oracle~$Y$. Now, let $L$ be a computable set that cannot be computed in time $2^{f(n+1)}$. Let $S$ be the sub-f-tree of~$T$ defined by $S(\sigma)=T(\sigma \oplus L)$ (where $\sigma \oplus L = \sigma(0)L(0) \ldots \sigma(k-1)L(k-1)$ when $k=|\sigma|$). The paths of $S$ are exactly the sets of the form $T(X \oplus L)$ for some~$X$. Each of them computes $L$ in time $f(n+1)$ by definition of~$f$, which is exponentially faster than any procedure computing~$L$ without oracle by our assumption on~$L$. 
\end{proof}

\begin{proof}[Proof of Theorem~\ref{thm:minimal-bound-no-lfsst}]
Let $T$ be a total computable f-tree and $\Phi_e$ a functional. As we explained earlier, the usual construction of a minimal degree shows that there is $S \preccurlyeq T$ which is either $e$-consistent or $e$-splitting. In the case $S$ is $e$-consistent, we are satisfied (this guarantees $\Phi_e^A$ to be either partial or computable). If it is $e$-splitting, we further refine~$S$ as follows. Since $S$ is $e$-splitting, we consider the total computable f-tree $S'$ corresponding to the image of $S$ by $\Phi_e$: $S'(\sigma)=\Phi_e^{S(\sigma)}$ (this is indeed an f-tree precisely because $S$ is $e$-splitting). By the previous lemma, there is a total computable $S'' \preccurlyeq S'$ none of whose paths is low for speed. Now the pullback $T' = \Phi_e^{-1}(S'')$ is a total computable f-subtree of~$T$, which forces $\Phi_e^A$ to not be low for speed. 

Thus we can force for all~$e$ that $\Phi_e^A$ is partial or not low for speed, and force $A$ to be of minimal degree and be non-computable as usual. 
\end{proof}

We do not know whether or not \emph{all} sets of minimal degree are in fact non-low for speed. \\

\begin{remark}
Another way to prove Theorem~\ref{thm:minimal-bound-no-lfsst} is to use Kumabe ans Lewis's theorem that there exists a minimal degree of DNC degree~\cite{KumabeL2009}, and apply Theorem~\ref{thm:dnc-not-lfs}, but our proof is more informative, as it also shows that lowness for speed is a generic notion for forcing with total with computable f-trees. 
\end{remark}


\vspace{8mm}

\noindent \textbf{Acknowledgements.} This paper grew from interactions between complexity theory and classical computability theory originating from the 2012 Dagstuhl Seminar ``Computability, Complexity and Randomness'' (Seminar 12012). Bienvenu acknowledges support of ANR-15-CE40-0016-01 RaCAF grant, Downey thanks the Marsden Fund of New Zealand, and the LIRMM (University of Montpellier) where this research was undertaken.

\bibliographystyle{alpha}
\bibliography{lfs-biblio}

\newcommand{\etalchar}[1]{$^{#1}$}
\begin{thebibliography}{CDE{\etalchar{+}}14}

\bibitem[ABK06]{AllenderBK2006}
Eric Allender, Harry Buhrman, and Michal Kouck{\'y}.
\newblock What can be efficiently reduced to the {K}olmogorov-random strings?
\newblock {\em Annals of Pure and Applied Logic}, 138:2--19, 2006.

\bibitem[AFG13]{AllenderFG2013}
Eric Allender, Luke Friedman, and William~I. Gasarch.
\newblock Limits on the computational power of random strings.
\newblock {\em Information and Computation}, 222:80--92, 2013.

\bibitem[Bay12]{Bayer-PhD}
Robertson Bayer.
\newblock {\em Lowness For Computational Speed}.
\newblock PhD thesis, University of California Berkeley, 2012.

\bibitem[BGS75]{BakerGS1975}
Theodore Baker, John Gill, and Robert Solovay.
\newblock Relativizations of the {$\mathcal{P} = ?\mathcal{NP}$} question.
\newblock {\em SIAM Journal on Computing}, 4(4):431--442, 1975.

\bibitem[Bie08]{Bienvenu2008-phd}
Laurent Bienvenu.
\newblock {\em Game-theoretic characterizations of randomness: unpredictability
  and stochasticity}.
\newblock PhD thesis, Universit{\'e} de Provence, 2008.
\newblock https://tel.archives-ouvertes.fr/tel-00332425v2.

\bibitem[Blu71]{Blum1971}
Manuel Blum.
\newblock On effective procedures for speeding up algorithms.
\newblock {\em Journal of the ACM}, 18(290-305), 1971.

\bibitem[CDE{\etalchar{+}}14]{CaiDELM2014}
Mingzhong Cai, Rodney Downey, Rachel Epstein, Steffen Lempp, and Joseph Miller.
\newblock Random strings and tt-degrees of {T}uring complete c.e. sets.
\newblock {\em Logical Methods in Computer Science}, 10(3), 2014.

\bibitem[DH10]{DowneyH2010}
Rodney Downey and Denis Hirschfeldt.
\newblock {\em Algorithmic randomness and complexity}.
\newblock Theory and Applications of Computability. Springer, 2010.

\bibitem[Hau86]{Haught1986}
Christine~A. Haught.
\newblock The degrees below a 1-generic degree {$<0'$}.
\newblock {\em Journal of Symbolic Logic}, 51, 1986.

\bibitem[Joc80]{Jockusch1980}
Carl Jockusch.
\newblock Degrees of generic sets.
\newblock In Frank Drake and Stanley~S. Wainer, editors, {\em Recursion theory:
  its generalizations and applications}, number~45 in London Mathematical
  Society Lecture Note Series, pages 110--139. Cambridge Unversity Press, 1980.

\bibitem[Joc89]{Jockusch1989}
Carl Jockusch.
\newblock {\em Degrees of functions with no fixed points}, pages 191--201.
\newblock North-Holland, Amsterdam, 1989.

\bibitem[Kau91]{Kautz1991}
Steven~M. Kautz.
\newblock {\em Degrees of random sets}.
\newblock PhD thesis, Cornell University, 1991.

\bibitem[KL09]{KumabeL2009}
Masahiro Kumabe and Andrew~{E. M.} Lewis.
\newblock A fixed-point-free minimal degree.
\newblock {\em Journal of the London Mathematical Society}, 80(3):785--797,
  2009.

\bibitem[Koz06]{Kozen2006}
Dexter Kozen.
\newblock {\em Theory of Computation}.
\newblock Springer, New York, 2006.

\bibitem[Ku{\v c}85]{Kucera1985}
Antonin Ku{\v c}era.
\newblock Measure, {$\Pi^0_1$} classes, and complete extensions of {PA}.
\newblock {\em Lecture Notes in Mathematics}, 1141:245--259, 1985.

\bibitem[Kur81]{Kurtz1981}
Stuart Kurtz.
\newblock {\em Randomness and genericity in the degrees of unsolvability}.
\newblock Ph{D} dissertation, University of Illinois at Urbana, 1981.

\bibitem[Nie09]{Nies2009}
Andr{\'e} Nies.
\newblock {\em Computability and randomness}.
\newblock Oxford Logic Guides. Oxford University Press, 2009.

\bibitem[NST05]{NiesST2005}
Andr{\'e} Nies, Frank Stephan, and Sebastiaan Terwijn.
\newblock Randomness, relativization and {T}uring degrees.
\newblock {\em Journal of Symbolic Logic}, 70:515--535, 2005.

\bibitem[Sac63]{Sacks1963b}
Gerald Sacks.
\newblock On the degrees less than {0'}.
\newblock {\em Annals of Mathematics}, 77:211--231, 1963.

\bibitem[Soa16]{Soare2016}
Robert Soare.
\newblock {\em Turing Computability: Theory and Applications}.
\newblock Theory and Applications of Computability. Springer, 2016.

\end{thebibliography}


\end{document}